\documentstyle[epsfig]{aipproc}

\newcommand{\gsim}{\buildrel>\over{_\sim}}
\newcommand{\lsim}{\buildrel<\over{_\sim}}

\begin{document}
\begin{flushright}
KEK-TH-739     \\
UT-ICEPP 01-01 \\
January 2001   \\
\end{flushright}

\title{Extracting SUSY Parameters from the Higgs Boson Properties
\footnote{Talk given by S.Kiyoura at the Linear Collider Workshop 2000,
          October 24-28, 2000, Fermilab.}}
\author{\footnote{shingo.kiyoura@kek.jp}
        Shingo Kiyoura${}^{a,b}$ and
        \footnote{yasuhiro.okada@kek.jp}
        Yasuhiro Okada${}^a$}
\address{${}^a$ Theory Group, KEK, Tsukuba, Ibaraki, 305-0801, Japan \\
${}^b$ ICEPP, University of Tokyo, Tokyo 113-0033, Japan}
\maketitle
\begin{abstract}
We calculate the ratio of the two branching ratios,
Br($h \rightarrow b\overline{b})$ and Br($h \rightarrow c\overline{c}$)
$+$ Br($h \rightarrow gg$), in the minimal supersymmetric standard model
taking into account the SUSY-loop corrections to the Higgs sector and
the $hb\overline{b}$ vertex.
We show that the heavy Higgs mass can be extracted from the ratio, almost
independently of other SUSY parameters, in the region of
$\tan\beta \lsim$ 10.
\end{abstract}
It was pointed out that the CP-odd Higgs mass $m_A$ can be determined
by the measurements of the lightest Higgs decay branching
ratios in the minimal supersymmetric standard model (MSSM)\cite{KOTrbr}.
Since there is the region of moderate $\tan\beta$
where LHC may not be able to detect the CP-odd Higgs boson,
$m_A \gsim 200$GeV\cite{was},
it is important to study such an option to indirectly constrain $m_A$.
Ref.\cite{KOTrbr} showed that we could extract $m_A$ from the double
ratio,
\begin{eqnarray*}
  R_{br} \equiv \frac{Br(h \rightarrow c \overline{c}) +
                  Br(h \rightarrow gg)}
                 {Br(h \rightarrow b \overline{b})},
\end{eqnarray*}
taking into account the SUSY-loop corrections to the Higgs sector.

Since the dependence of the $h \rightarrow b\overline{b}$ and
$h \rightarrow c\overline{c}$ branching ratios on the mixing 
angles $\alpha$ and $\beta$ of the Higgs sector is given by
$Br(h \rightarrow b \overline{b}) \propto
    \frac{\sin^2 \alpha}{\cos^2 \beta}$ and
$Br(h \rightarrow c \overline{c}) \propto
    \frac{\cos^2 \alpha}{\sin^2 \beta}$,
the double ratio between $Br(h \rightarrow b \overline{b})$ and 
$Br(h \rightarrow c \overline{c})$ becomes
$ R_c \equiv Br(h \rightarrow c \overline{c})/Br(h \rightarrow b \overline{b})
  \propto 1 / (\tan \alpha \tan \beta)^2$,
where the scalar-top-quark corrections to the Higgs sector are implicit
in $\tan\alpha$ and $\tan\beta$.
When the CP-odd Higgs boson is heavy, $m_A \gg m_h \sim m_Z$,
and the left-right mixing of the scalar top quarks is small,
the double ratio $R_c$ is approximately proportional to
$(m_h^2-m_A^2)^2/(m_Z^2+m_A^2)^2$.
From this expression, you can see that, once we measure the lightest Higgs
mass $m_h$, the double ratio $R_c$ determines $m_A$.
Furthermore, $Br(h \rightarrow gg)$, dominantly induced by the top quark
exchange, has the same dependence on the mixing angles $\alpha$ and $\beta$
as the $Br(h \rightarrow c\overline{c})$, and therefore we can expect
the double ratio $R_{br}$ determines the mass scale of the CP-odd Higgs
boson.

In the above discussion, we assume that Higgs-fermion Yukawa couplings
are the same as those in the type-II two Higgs doublet model (THDM).
Recently, the SUSY-loop corrections to the $hb\overline{b}$ coupling constant
have been studied by many authors\cite{hbb,hH}.
The one-loop-level coupling of a bottom quark to the neutral Higgs fields
is given by
\begin{eqnarray}
  {\cal L} &=& f_b \overline{b}_L b_R H_d^0 +
             \epsilon_b f_b \overline{b}_L b_R H_u^{0*} + h.c.,
             \label{hc} \\
  && \epsilon_b \equiv
     \frac{2 \alpha_S}{3\pi} \mu M_3
         f(M_3^2,m_{\tilde{b}_L}^2,m_{\tilde{b}_R}^2) +
     \frac{f_t^2}{16\pi^2} \mu A_t 
         f(\mu^2,m_{\tilde{t}_L}^2,m_{\tilde{t}_R}^2), \label{eps_b}\\
  && f(m_1^2,m_2^2,m_3^2) \equiv \frac{1}{m_3^2}
        \left[ \frac{x\log x}{1-x} - \frac{y\log y}{1-y} \right]
        \frac{1}{x-y}, ~~
     x \equiv \frac{m_1^2}{m_3^2}, ~~
     y \equiv \frac{m_2^2}{m_3^2}
\end{eqnarray}
where $\epsilon_b$ is induced by the gluino- and Higgsino-exchange diagrams.
In Eq.(\ref{hc}) the second term proportional to $\epsilon_b$
is absent at the tree level in the MSSM (and also in the type-II THDM).
The b-quark mass and the $hb\overline{b}$ coupling constant are
then expressed as,
\begin{eqnarray*}
  m_b &=& f_b v_d + f_b \epsilon_b v_u =
          f_b v \cos \beta \times (1 + \epsilon_b \tan \beta), ~~
  {\cal L}_{hbb} = - \frac{m_b \sin \alpha}{v \cos \beta} \times
      \left[
        \frac{1-\epsilon_b / \tan \alpha}{1+\epsilon_b \tan \beta}
      \right].
\end{eqnarray*}
We can see that the effect of the $\epsilon_b$ on the b-quark mass and
the $hb\overline{b}$ coupling constant becomes significant
when $\tan\beta$ is large. In this case the effective theory below
the SUSY-breaking scale becomes the general THDM.

The double ratios between the Higgs decay branching ratios are
proportional to the following expressions:
\begin{eqnarray*}
  R_{br} &\propto&
    \frac{1}{(\tan \alpha \tan\beta)^2}
    \left[
      \frac{(1+\epsilon_b \tan \beta)}{(1-\epsilon_b / \tan \alpha)}
    \right]^2 \label{Rccbb}, \\
  R_\tau &\equiv&
    \frac{Br(h\rightarrow \tau^+\tau^-)}{Br(h\rightarrow b\overline{b})}
    \propto
    \left[
      \frac{(1+\epsilon_b \tan \beta)}{(1-\epsilon_b / \tan \alpha)}
    \right]^2.
\end{eqnarray*}
When $\tan\beta$ is large, the effect of $\epsilon_b$ on the double ratios,
$R_{br}$ and $R_\tau$, becomes significant whereas the ratio $R_{br}/R_\tau$
is $\epsilon_b$ independent. In addition, if the stop mixing parameter
$A_t$ is large, the SUSY-loop corrections to the Higgs sector modify
the approximate relation,
$1/(\tan \alpha \tan \beta)^2 \sim (m_h^2-m_A^2)^2/(m_Z^2+m_A^2)^2$\cite{hH}.


We first consider the uncertainties of the decay double ratios 
for the standard model (SM) Higgs boson due to the SM input parameters;
the strong coupling
constant $\hat{\alpha}_S(m_Z)$, and the $\overline{\mbox{MS}}$ running quark
masses of the bottom and charm quarks, $\hat{m}_b(m_b)$ and
$\hat{m}_c(m_c)$.
We assume the following center values and errors;
$\hat{\alpha}_S(m_Z)=0.1181 \pm 0.002,
\hat{m}_b(m_b)=4.20 \pm 0.13 ~\mbox{GeV} ~(\pm3\%),
\hat{m}_c(m_c)=1.25 \pm 0.06 ~\mbox{GeV} ~(\pm5\%).$
For the $h \rightarrow q\overline{q}$ and $h \rightarrow gg$ partial widths,
we used the same formulas as in Ref.\cite{KOTrbr}.
In the second column in Table \ref{sm_error}, we show the SM theoretical
uncertainties for the ratios obtained by varying the input parameters
in the above range. In the third column, the expected statistical errors of
the double ratios are given for the integrated luminosity of 100 fb${}^{-1}$
and 500 fb${}^{-1}$ obtained by scaling the results of Ref.\cite{Nakamura}.
Totals of the theoretical and experimental errors for the three
double ratios are shown in the forth column for the integrated luminosity of
100 fb${}^{-1}$ and 500 fb${}^{-1}$ respectively. Assuming the integrated
luminosity of 500 fb${}^{-1}$, total errors for the three double ratios are
about 10\%.
\begin{table}[h]
\caption{The errors of the double ratios of the Standard Model Higgs
boson due to the theoretical uncertainties of input parameters and
experimental errors (\%).}
\label{sm_error}
\begin{tabular}{l|ccc}
%
%
 & Theoretical error & Experimental error
 & Total error \\
 &             & (100/500) fb${}^{-1}$
 & (100/500) fb${}^{-1}$ \\
\tableline
%
%
$R_{br}$ = $\frac{Br(h->c\overline{c})+Br(h->gg)}{Br(h->b\overline{b})}$ &
  8.6 & 15.0/6.7 & 17.3/10.9 \\
%
%
$R_\tau$ = $\frac{Br(h->\tau^+ \tau^-)}{Br(h->b\overline{b})}$ &
  7.6 & 10.0/4.5 & 12.6/8.8 \\
%
%
$R_{br}/R_\tau$ = $\frac{Br(h->c\overline{c})+Br(h->gg)}{Br(h->\tau^+ \tau^-)}$
  & 4.1 & 19.5/8.7 & 19.9/9.6 \\
\end{tabular}
\end{table}


We present the double ratios as functions of $M_A$ for typical SUSY
parameters. For this purpose we introduce the SUSY-breaking scale
$M_S$ and set the soft masses for squarks as
$m_{\tilde{q}_L}^2 = m_{\tilde{t}_R}^2 = m_{\tilde{b}_R}^2 = M^2_S$,
and the squark mixing parameters as $A_t = X_t M_S$ and $A_b=0$,
where $X_t$ is a dimensionless parameter.
Then we can calculate the Higgs masses
$m_{h,H}$, the mixing angle $\alpha$ and the radiatively induced coupling
$\epsilon_b$ in Eq.(\ref{eps_b}) from the parameter set
($\tan\beta$, $M_A$, $M_S$, $X_t$, $M_3$, $\mu$).
In our numerical calculation, we solve the renormalization group equations
of the Higgs sector given in Ref.\cite{Haber} to obtain the Higgs boson
masses and the mixing angle $\alpha$.

\begin{figure}[h]
\centerline{\epsfig{file=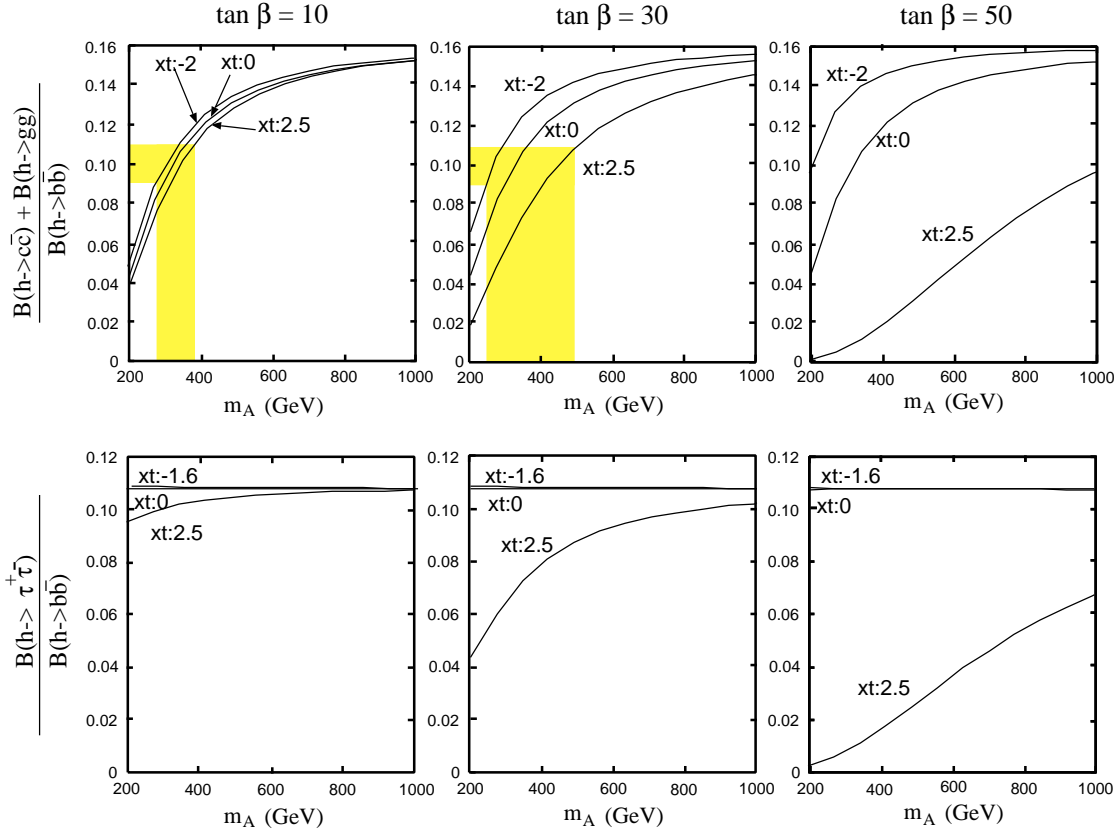, height=11cm}}
\caption{$R_{br}$ (upper row) and $R_\tau$ (lower row) of the 120GeV
lightest neutral Higgs boson for $\tan\beta=10,30,50$.
%
}
\label{plot}
\end{figure}

In Fig.\ref{plot} we plot the double ratios
$R_{br}$ and $R_\tau$ in the upper and lower rows respectively for
$\tan\beta=$ 10, 30, and 50 as functions of $M_A$.
We set ($M_3$, $\mu$) to (300GeV, 300GeV) 
and solve $M_S$ such that the lightest Higgs mass $m_{h^0}$ becomes
120GeV for each $M_A$ with fixing the ratio $X_t$.
For the present parameter set, $R_{br}$ has the maximum (minimum) value
when $X_t$ is set to $-2 ~(2.5)$ and $R_\tau$ has the maximum (minimum)
value at $X_t= -1.6 ~(2.5)$ in the range of $-2.5 \leq X_t \leq 2.5$.
The dependence of $X_t$ comes from both the induced coupling
$\epsilon_b$ and the mixing angle $\alpha$. The shaded regions in the graphs
for $\tan\beta=10, 30$ show the constrained region of
the CP-odd Higgs mass assuming that $R_{br}$ were determined with the 10\%
accuracy.
We can see that $M_A$ is well (weakly) constrained by $R_{br}$, 
when $\tan\beta$=10 (30). When $R_{br}$ receives the significant
corrections, $R_{\tau}$ also receives corrections of the similar
magnitude. For moderate $\tan\beta\sim$10, both the ratios $R_{br}$
and $R_{br}/R_\tau$ are approximately proportional to
$(m_h^2-m_A^2)^2/(m_Z^2+m_A^2)^2$.

To summarize, we estimated the theoretical uncertainties due to
the SUSY-loop corrections on the double ratios between the branching
fractions of the lightest Higgs boson. In the region of moderate
$\tan\beta \sim 10$, where LHC will not be able to detect
the CP-odd Higgs boson of $m_A \gsim 200$GeV\cite{was}, we can constrain
the range of the CP-odd Higgs mass from the ratios, $R_{br}$ and
$R_{br}/R_\tau$.
On the other hand, if LHC detects the CP-odd Higgs boson, we may be
able to obtain information on the SUSY sector by the branching ratios of 
the Higgs boson at the future LC. \\

YO  was supported by the Grant-in-Aid of the Ministry of Education,
Science, Sports and Culture, Government of Japan (No.\ 09640381),
Priority area ``Supersymmetry and Unified Theory of Elementary Particles''
(No.\ 707), and ``Physics of CP Violation'' (No.\ 09246105).

\end{document}